\documentclass[10pt,twocolumn,letterpaper]{article}

\usepackage[pagenumbers]{cvpr} 

\definecolor{cvprblue}{rgb}{0.21,0.49,0.74}
\usepackage[pagebackref,breaklinks,colorlinks,allcolors=cvprblue]{hyperref}
\usepackage{overpic}
\usepackage{adjustbox}
\usepackage{multirow}
\def\paperID{6437} 
\def\confName{CVPR}
\def\confYear{2025}
\usepackage{pifont}
\usepackage{arydshln}
\usepackage{dsfont}
\newcommand{\cmark}{\ding{51}}
\newcommand{\xmark}{\ding{55}}
\title{Towards Effective User Attribution for Latent Diffusion Models via Watermark-Informed Blending}

\author{
    Yongyang Pan$^{1}$, Xiaohong Liu$^{1}$\thanks{Corresponding author.}, Siqi Luo$^{1}$, Yi Xin$^{2}$, Xiao Guo$^{3}$,\\ Xiaoming Liu$^{3}$, Xiongkuo Min$^{1}$, Guangtao Zhai$^{1}$\\
    $^{1}$Shanghai Jiao Tong University, $^{2}$Nanjing University, $^{3}$Michigan State University\\
    {\tt\small \{panyongyang, xiaohongliu, siqiluo647, minxiongkuo, zhaiguangtao\}@sjtu.edu.cn, }\\
    {\tt\small xinyi@smail.nju.edu.cn,}
    {\tt\small \{guoxia11, liuxm\}@msu.edu}
}
\begin{document}
\maketitle
\begin{abstract}
Rapid advancements in multimodal large language models have enabled the creation of hyper-realistic images from textual descriptions. However, these advancements also raise significant concerns about unauthorized use, which hinders their broader distribution. Traditional watermarking methods often require complex integration or degrade image quality. To address these challenges, we introduce a novel framework \underline{T}owards \underline{E}ffective user \underline{A}ttribution for latent diffusion models via \underline{W}atermark-\underline{I}nformed \underline{B}lending (TEAWIB). TEAWIB incorporates a unique ready-to-use configuration approach that allows seamless integration of user-specific watermarks into generative models. This approach ensures that each user can directly apply a pre-configured set of parameters to the model without altering the original model parameters or compromising image quality. Additionally, noise and augmentation operations are embedded at the pixel level to further secure and stabilize watermarked images. Extensive experiments validate the effectiveness of TEAWIB, showcasing the state-of-the-art performance in perceptual quality and attribution accuracy. \textit{The code will be released upon publication.}
\end{abstract}    
\section{Introduction}
The advent of text-to-image models has revolutionized the creation of photorealistic images by transforming descriptive text into visually accurate representations, significantly advancing the capabilities of image synthesis~\cite{ref1, ref4,  yang2024diffstegauniversaltrainingfreecoverless}. Contemporary Latent Diffusion Models (LDMs), including Stable Diffusion~\cite{ref3} and DALL·E 2~\cite{ref2}, exhibit an impressive capacity to generate a wide range of novel images across diverse scenes and contents. While these advancements represent a substantial leap in AI-Generated Content (AIGC), they simultaneously raise concerns about the potential misuse of these models~\cite{ref5}. 
\begin{figure}[t]
    \centering
    \includegraphics[width=0.48\textwidth]{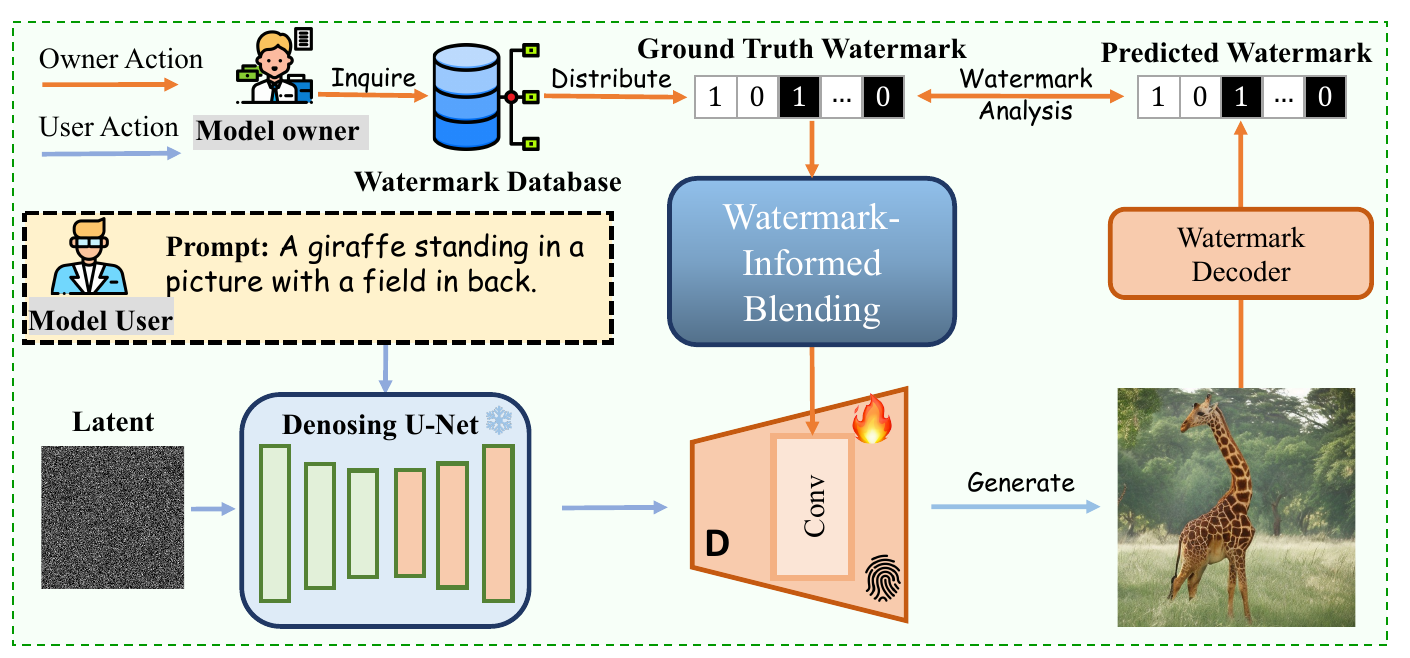}
    \caption{\textbf{Workflow of TEAWIB.} During inference, the fingerprinted model (\textit{i.e.}, $\mathcal{D}$) is employed by the model user to generate images embedded with an invisible watermark. The model owner can then utilize the watermark decoder to identify watermarks within questionable images and trace their origins back to the respective model users.}
    \label{fig:pipeline}
\end{figure}

\par
Recent advances have significantly enhanced watermark embedding techniques in images \cite{guo2023hierarchicalfinegrainedimageforgery, liu2022psccnetprogressivespatiochannelcorrelation, ref27, luo2020distortionagnosticdeepwatermarking, kishore2021fixed, fu2023rawiwrawimagewatermarking, wu2023cheapfakedetectionllmusing}.
However, these methods are typically applied in a post-generation manner, making the watermarks vulnerable to manipulation, especially when models like Stable Diffusion (SD)~\cite{ref3} are leaked or open-sourced. In these cases, a simple code modification can bypass the watermarking process. To address this vulnerability, the integration of watermark generation during the image creation process has emerged as a recent trend. For instance, Stable Signature~\cite{ref21} has developed an ad-hoc watermarking method that fine-tunes the decoder with a pre-defined watermark. Additionally, WOUAF~\cite{ref22} advances this approach by modifying decoder parameters during the image generation phase, thereby embedding watermarks more securely and supporting scalable watermarking.

Despite these advancements, integrated watermarking methods currently face two limitations. First, Stable Signature~\cite{ref21} requires retraining the model with each watermark update and only supports a single designated watermark, leading to training costs that scale linearly with the number of users. Second, methods like WOUAF, which modulate model parameters, can degrade image quality and increase vulnerability to post-processing attacks. Additionally, such methods require modifications to the base model parameters, further compromising the quality of the generated images.

To address these challenges, we introduce TEAWIB, a novel framework that adopts a proposed watermark-informed blending strategy, which comprises two components: Dynamic Watermark Blending (DWB) and Image Quality Preservation (IQP). The DWB module includes watermark-specific weights, enabling the model to support scalable watermarks and avoiding the retraining procedure. The IQP module uses noise and augmentation operations to subtly embed user-specific information while preserving the high quality of generated images. Consequently, our method achieves seamless integration of watermarking into generative models without compromising image quality.
TEAWIB distinguishes itself by enabling the effective management of a substantial user population while maintaining superior image quality over existing integrated watermark embedding methods. 

The workflow of TEAWIB is illustrated in Fig.~\ref{fig:pipeline}. Our framework unfolds through three key steps: \textbf{(1)} The model owner distributes a user-specific watermark and then embeds it into the network as its fingerprint.
\textbf{(2)} the model owner modifies pre-trained parameters of the SD decoder. This integration process does not alter the original model parameters but instead enriches them with user-specific watermark information. The integration employs a dynamic blending factor that optimally balances the original pre-trained weights with the new user-specific adjustments. 
Additionally, noise and augmentation operations are integrated at the pixel level to embed user-specific information subtly, enhancing the robustness of watermarked images against post-processing manipulations. 
\textbf{(3)} The model user generates images using the fingerprinted decoder, seamlessly embedding invisible watermarks in generated images. This allows the model owner to detect watermarks from potentially unauthorized images.

In summary, our key contributions are:



\begin{itemize}

    
    \item We introduce TEAWIB, a novel framework that dynamically and seamlessly integrates watermarking into generative models without compromising the image quality. More importantly, TEAWIB incorporates a ``Ready-to-Use" manner, facilitating straightforward integration into existing models without the need for retraining, making it highly accessible and practical in real-world applications.
    
    \item To realize TEAWIB, we propose two key modules: Dynamic Watermark Blending (DWB) and Image Quality Preservation (IQP). DWB dynamically adjusts the balance between pre-trained and watermark-specific weights, optimizing watermark robustness without compromising image quality. Concurrently, IQP employs noise and augmentation operations to subtly embed user-specific information, thus maintaining the high-quality appearance of images and ensuring both the effectiveness and invisibility of watermarks.

    \item Comprehensive experiments on the MS-COCO validation dataset demonstrate the superiority of TEAWIB in both visual quality and attribution accuracy, establishing its position as a state-of-the-art and scalable watermark embedding method.
\end{itemize}

    
\section{Related Work}

\subsection{Traditional Watermarking Methods}
Traditional watermarking techniques~\cite{ref9, ref10, ref11, ref12} focus on embedding messages into the frequency domain of images, using methods like the Discrete Cosine Transform (DCT) and Discrete Wavelet Transform (DWT). These approaches typically modify specific frequency coefficients to embed watermarks but are applied post-hoc, making them easy for malicious users to bypass. In contrast, our ad-hoc watermarking method integrates watermark embedding directly into the image generation process, making it more resilient to tampering. While~\cite{ref20} inserts a predefined tree-ring watermark into the Fourier space of the latent vector, it is restricted to the DDIM scheduler. Our method, however, is robust to various attacks and compatible with multiple schedulers, providing greater flexibility and security.

\subsection{LDM-oriented Watermarking}
LDM-oriented watermarking modifies either the initialization of latent vectors or the decoder within LDMs. ~\cite{ref19} freezes the decoder and trains the initialization of latent vectors to embed watermarks, preserving perceptual quality. ~\cite{ref21} fine-tunes the decoder of LDM with a user-specific watermark, achieving notable results. However, this method is restricted to a single watermark and requires re-training for any change. ~\cite{ref22} applies weight modulation to the convolutional layers in the decoder of SD, facilitating the distribution of multiple watermarks. While this method attains satisfactory attribution accuracy, it compromises image quality. Our proposed method addresses this limitation by introducing a watermark-informed blending strategy.

This strategy achieves a balance between modified weights and pre-trained weights, resulting in the generation of high-quality images. It also integrates seamlessly into the model through a Ready-to-Use configuration, facilitating easy adoption without the need of complex modifications.


\section{Method}\label{sec:method}
\subsection{Preliminaries}
Our TEAWIB is primarily designed for Stable Diffusion models (SD). For clarity, SD consists of three principal components: the encoder $\mathcal{E}$, which converts an image $I$ into a latent vector $z$; the diffusion model $\epsilon_{\theta}$, built upon the U-Net architecture and augmented by a cross-attention mechanism for textual manipulation; and the decoder $\mathcal{D}$, which reconstructs an image $\hat{I}=\mathcal{D}(z)$ from the latent vector $z$. Both the encoder and the decoder are extensively trained on a large dataset to generate realistic images.

\begin{figure*}[tbp]
    \centering
    \begin{overpic}[width=1.01\linewidth]{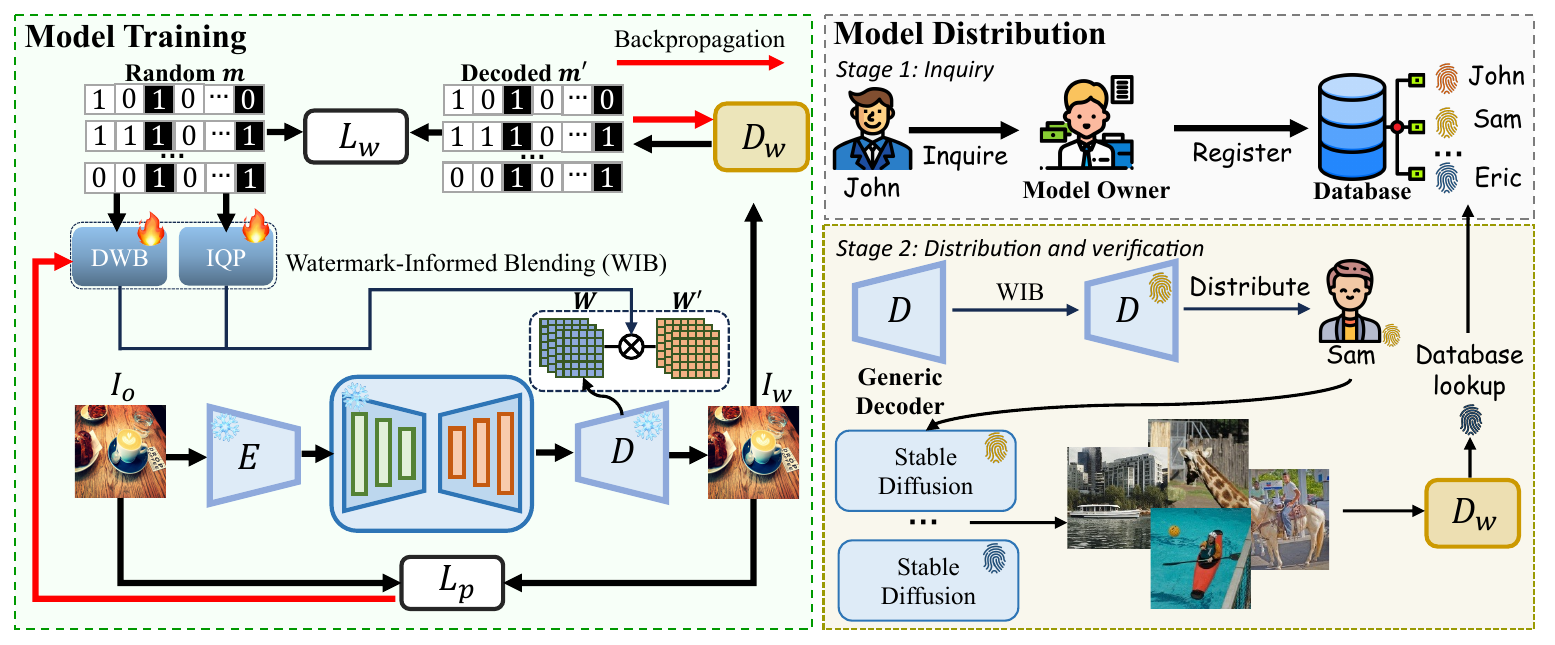}
    \end{overpic}
    \caption{\textbf{Overview of TEAWIB.} The workflow is bifurcated into two primary phases: (1) \textit{model training} and (2) \textit{model distribution}. During the training phase, the watermark decoder and the diffusion decoder are concurrently trained using a variety of randomly generated watermarks. The model distribution phase consists of two stages. Stage 1: The model user (e.g., John) requests access from the model owner, who assigns a unique watermark for the user and registers it in the database. Stage 2: For each authorized model user (e.g., Sam), the model owner selects a user-specific watermark from the database and integrates it with the generic decoder using the WIB method, distributing the watermarked decoder to the user. Model users can then employ these text-to-image models for image generation, with embedded invisible watermarks. For verification, the model owner can decode the watermark from any misused image and match it with the database to identify the specific user.}
    \label{fig:model overview}
\end{figure*}

\subsection{Overview}
As depicted in Fig. \ref{fig:model overview}, the proposed method is divided into two main stages: (1) model training and (2) model distribution. Modifications are focused solely on the decoder $\mathcal{D}$, with the diffusion process remaining unchanged, ensuring compatibility across a variety of generative tasks. The model distributor uses a centralized database to manage the keys assigned to each model user. The model owner is tasked with the creation and maintenance of this database, in which each record links a user ID to a specific watermark string $m \sim \mathbf{Ber}(0.5)^{d_w}$, of length $d_w$.

In model training, a designated watermark message $m$ is selected. This watermark is subsequently transformed into a fingerprint $r_w$ through a watermark mapping network $E_w$ (built upon two fully connected layers). Subsequently, the fingerprint $r_w$ is leveraged by the proposed watermark-informed blending that embeds the watermark information to get fingerprinted weights $\textbf{W}^{\prime}$. 
During the training phase, the decoder $\mathcal{D}$ utilizes both $z$ and $r_w$ as inputs to generate the watermarked image $I_w$. A watermark decoder $\mathcal{D}_w$ is then employed to extract the watermark message $m$ from this image. It is worth noting that during training, the pre-trained weights in the convolutional layers are frozen, w with updates applied only to the parameters in the  IQP, DWB modules and $\mathcal{D}_w$.

The model distribution phase consists of two stages. In the first stage, the model user requests access from the model owner, who registers the user in the database and assigns a unique watermark, provided the user is authorized. In the second stage, the model owner fingerprints the trained generic decoder with the user-specific watermark and distributes the fingerprinted model to the respective user. It is worth noting that the model has a Ready-to-Use configuration, allowing the owner to easily incorporate the watermark by simply providing it to the generic model. Model users can then utilize fingerprinted text-to-image models for image generation. For verification, the model owner can extract the watermark from any misused image and cross-reference it with the database to identify the specific user.

\subsection{Watermark-Informed Blending}
The proposed watermark-informed blending optimally integrates user-specific watermarks into the model decoder during image generation, ensuring merged-in watermark generation while preserving quality. 
It comprises two key components: Dynamic Watermark Blending (DWB) and Image Quality Preservation (IQP). 
The DWB adjusts the blending of original and watermark-specific weights dynamically, ensuring robust watermark embedding with minimal impact on image quality. Simultaneously, the IQP maintains the high quality of images through noise and augmentation operations, enhancing the robustness of the watermark to digital manipulations. Together, these two components ensure that watermarks are effective yet imperceptible, preserving the aesthetic and utility of the generated images.
\subsubsection{Dynamic Watermark Blending.}
Previous methods \cite{ref22} directly modulate the kernel weights and apply them in convolution:
\begin{equation}\label{eq: conv moulation}
    y = f_c(\textbf{W}^{\prime},x).
\end{equation}
Here, where $y$ denotes the output feature map, $\textbf{W}^{\prime}$ and $x$ denote the fingerprinted weight and input, respectively. $f_c$ represents the convolutional operation.

In Eq. \ref{eq: conv moulation}, pre-trained weights are modified during fine-tuning. This can lead to issues such as domain shift if the training dataset significantly differs from the original training dataset. 
Additionally, neglecting the pre-trained weights could potentially result in the loss of valuable information from the initial large-scale training.
Pre-trained weights are responsible for generating hyper-realistic images. 
Additionally, embedding watermark information into the model’s parameters is key to implementing ad-hoc watermarking in the Stable Diffusion model. 
Considering these two factors, we propose the watermark-informed blending that introduces a blending operation into the convolutional process, incorporating the influence of both the original weights and the fingerprinted weights.
Inspired by alpha digital image compositing, the final representation of this submodule is formulated as:
\begin{equation}\label{eq:method overall}
y_d = \alpha\cdot f_c(\textbf{W}^{\prime},x) + (1 - \alpha)\cdot f_c(\textbf{W},x),
\end{equation}
where $y_d$ denotes the blended output feature map, $\textbf{W}$ represents the pre-trained weights for the convolutional layer, and $\alpha$ is a trainable blending factor. During training, $\alpha$ is optimized based on watermark detection accuracy and perceptual similarity between watermarked images and training images.

\noindent\underline{\textit{{Fingerprint Insertion.}}}
A fundamental aspect of our approach is the direct insertion of watermark messages into the model parameters. To integrate them seamlessly, we modulate the convolutional weights of the decoder $\mathcal{D}$ using our fingerprint $r_w$ (the encoded watermark). This technique is inspired by StyleGAN2 \cite{ref30}.

Given a convolutional kernel $\mathbf{W} \in \mathbb{R}^{i \times j \times k}$ at layer $l$, where $i$, $j$, and $k$ denote input channels, output channels, and kernel size, respectively. 
Initially, a scale factor $s$ and a bias term $b$ are computed. 
The scale $s$ is obtained by projecting the fingerprint $r_w$ through a Multi-Layer Perceptron (MLP). 
The bias term $b$ is calculated as $A_l(r_w)$ using an affine transformation layer $A_l$, where weights of $A_l$ are initialized to zero and the biases are initialized to one. Consequently, $A_l(r_w)$ initially equals a tensor with all ones. 
This initialization of the bias term provides a stable baseline, ensuring that the scaling of $\mathbf{W}$ is not too extreme at the beginning. 
The introduction of two learnable terms, $s$ and $b$, allows the model to better adapt to the data, leading to improved performance.

Subsequently, the channel-wise weight modulation is then executed according to:
\begin{equation}\label{eq:weight modulation}
    \mathbf{W}^{\prime} = \mathbf{W}\cdot (s + b), 
\end{equation}
where $\mathbf{W}$ and $\mathbf{W}^{\prime}$ denote pre-trained and modulated kernel weights, respectively. 
We conduct weight modulation for all convolutional layers with the same fingerprint $r_w$. $\mathbf{W}^{\prime}$ is not a saved parameter, instead our method calculate this weights at every convolutional layer.

\subsubsection{Image Quality Preservation.}
The IQP consists of two operations to preserve image quality and enhance model robustness. Its functionality can be summarized as follows: 
\begin{equation} y_i = aug_{t} + \lambda_n \cdot \mathcal{N}(0, \sigma^2), \end{equation}
where $y_i$ represents the combined result of the two operations: $aug_{t}$, which is the augmentation operation, and $\lambda_n \cdot \mathcal{N}(0, \sigma^2)$, which is the noise operation. $\lambda_n$ is a trainable parameter initialized as $0$ to control the magnitude of noise addition.

\noindent\underline{\textit{Noise Operation.}} To increase the stochastic variation of generated images and refine intricate details, we introduce a randomly generated Gaussian noise $\mathcal{N}(0, \sigma^2)$, which $\sigma$ represents the standard deviation and sets to $1$ following \cite{ref23}.

\noindent\underline{\textit{Augmentation Operation.}} Although the noise operation refines the details, it can degrade robustness against post-processing attacks. 
To address this issue, we propose an augmentation operation, denoted as $aug_{t}$. For layer $l$, we project the fingerprint $r_w$ through $M_l$ (built upon two fully connected layers), \textit{i.e.}, $aug_t = M_{l}(r_w)$. 
This augmentation step operates at the pixel level to embed watermark-related information. Thorough experiments validate that the combination of these two modules can enhance the robustness of the watermark while preserving image quality.
The final output $y$ of the Watermark-Informed Blending is formulated as follows:
\begin{equation}
    y = y_i + y_d.
\end{equation}
It incorporates the DWB and IQP to improve both image quality and robustness towards image attacks. 
\begin{table*}[htbp]
    \hspace*{-5mm}
    \centering
    \footnotesize
    \small
    \setlength{\tabcolsep}{8pt}
    \caption{
        Comparison of TEAWIB with other merged-in generation methods. Experiments are conducted on images of size $512\times 512$ with $48$-bit watermarks. PSNR and SSIM are measured between the generations of the original and watermarked decoders, while FID is computed between original images and watermarked text-generated images. Bit accuracy, the percentage of correctly decoded bits, evaluates the robustness of the embedded watermarks. The terms ``Crop'' and ``Brigh.'' represent image post-processing steps that crop the image to half of its original size and adjust the brightness by a factor of two. TEAWIB achieves superior performance in terms of generation quality and robustness while being seamlessly integrated into the generation process.
    }
    \label{tab:quality-watermarking} 
        \begin{adjustbox}{max width=\textwidth}
        \begin{tabular}{ c c @{\hspace{2pt}} c  *{2}{c} *{2}{c} *{1}{c} *{3}{c}}
        \toprule
       &\multirow{2}{*}{Model}  & \multirow{2}{*}{Scalable Watermark}          & \multirow{2}{*}{PSNR / SSIM $\uparrow$} & \multirow{2}{*}{$L_{\infty}$ $\downarrow$} & \multirow{2}{*}{LPIPS $\downarrow$} & \multirow{2}{*}{FID $\downarrow$} &\multicolumn{3}{c}{Bit accuracy $\uparrow$ on:} \\ 

    &                                         &                           &           &           &           &                                            &  None & Crop & Brigh. \\ \midrule
           & Stable Diffusion \cite{ref3}                        & $-$            &  $30.0$ / $0.89$ & $-$ & $-$ & $9.294$      & $-$ & $-$ & $-$ \\
           & Stable Signature \cite{ref21}                          & \xmark            &  $30.0$ / $0.89$ & $82.591$ & $0.0330$ & $9.376$      & $0.9918$ & $\textbf{0.9903}$ & $0.9519$ \\
            & WOUAF \cite{ref22}                            & \cmark        & $28.2$ / $0.82$ & $130.49$ & $0.0781$ & $18.996$
                & $0.9767$ & $0.5912$ & $0.6191$ \\
           & TEAWIB (Ours)                             & \cmark        & $\textbf{39.2}$ / $\textbf{0.98}$ & $\textbf{57.8}$ & $\textbf{0.0047}$ & $\textbf{9.223}$
                & $\textbf{0.9919}$ & $0.9820$  & $\textbf{0.9535}$ \\
            
        \bottomrule \vspace*{-0.2cm}
    \end{tabular}
     \end{adjustbox}
\end{table*}
\subsection{Loss Function}
In the training phase, our primary objectives are twofold: 1) accurate message extraction from watermarked images and 2) negligible effect of watermark insertion on the generated images. To achieve the former, we introduce the watermark extraction loss, denoted as $\mathcal{L}_w$, which also trains a pre-trained watermark extractor $\mathcal{D}_w$. The message extraction loss between watermark message $m$ and extracted message $m^\prime$ is defined as:
\begin{equation}\label{eq:binary cross entropy}
    \mathcal{L}_w = -\sum_{i=1}^{d_w} m_i\cdot \log\epsilon(m_i^\prime) + (1-m_i)\cdot\log(1-\epsilon(m_i^\prime)),
\end{equation}
where $\epsilon$ represents the Sigmoid function. For the latter objective, we utilize the perceptual loss $\mathcal{L}_p$. The loss is defined as:
\begin{equation} \mathcal{L}_p = \lambda_p\cdot\mathcal{L}_{v}(I_w, I_o) + \lambda_l \cdot \mathcal{L}_l(I_w, I_o),
\end{equation}
where $I_w$ and $I_o$ are watermarked images and images generated by the original decoder, respectively.
$\mathcal{L}_{v}$ refers to Watson-VGG loss \cite{ref31}, and $\mathcal{L}_{l}$ represents LPIPS-loss \cite{ref32}. $\lambda_p=0.2,\lambda_l=1$ are prescribed coefficients.
The total loss $\mathcal{L}$ is given by:
\begin{equation}
    \mathcal{L} = \lambda_w\mathcal{L}_w + \mathcal{L}_p,
\end{equation}
where $\lambda_w=1$ balances the watermark extraction loss and perceptual loss. The joint optimization of $\mathcal{D}$ and $\mathcal{D}_w$ facilitates the adaptation of $\mathcal{D}_w$ to watermark variations, enhancing its detection accuracy.

\section{Experiments} 
\subsection{Experimental Settings}
\subsubsection{{Datasets.}} The training and evaluation are conducted on the MS-COCO \cite{ref24} dataset of size $256\times 256$. In line with \cite{ref21}, $1,000$ text prompts from the MS-COCO validation dataset are selected to generate watermarked images.\par
\subsubsection{Diffusion Model Setup.} The adopted SD model in TEAWIB uses the same configuration as the repository\footnote{https://github.com/Stability-AI/stablediffusion}. Specifically, Stable Diffusion 2.0-base is chosen. For text-to-image generation, a guidance scale of 3.0 and 50 diffusion steps are used.\par
\subsubsection{Experimental Setup.} Drawing inspiration from the implementation in StyleGAN2-ADA \cite{ref25}, the watermark mapping network $E_w$ comprises two fully connected layers, each with a dimensionality matching that of the fingerprint $r_w$. For training, the AdamW optimizer \cite{ref26} is employed with a learning rate of $10^{-4}$. Additionally, the pre-trained watermark extractor from HiDDeN \cite{ref27} is leveraged.\par
\subsubsection{Evaluation Metrics.} 
To evaluate the quality of generated images, we use PSNR, SSIM \cite{ssim}, and LPIPS \cite{ref32} to measure the similarity between watermarked and original images. Furthermore, we quantify the realism and diversity of watermarked images via FID \cite{heusel2018gans}, and the $L_\infty$ norm is utilized to represent the maximum L-infinity norm between the watermarked and original images. The accuracy of watermark detection, denoted as  acc , is measured by Eq. \ref{eq:bit accuracy}. 
\begin{equation}\label{eq:bit accuracy}
   acc = \frac{1}{d_w} \sum_{i=1}^{d_w} \mathds{1}(m_i = m_i^\prime),
\end{equation}
where \( m^\prime = \mathcal{D}_w(I_w) \) represents the decoded watermark message from the generated image \( I_w \), and \( \mathds{1}(m_i = m_i^\prime) \) is an indicator function that equals 1 if \( m_i \) matches \( m_i^\prime \), and 0 otherwise.
 
\subsection{Experimental Results}
\subsubsection{Image quality.} 
\begin{figure*}[h]
\centering
\vspace*{0.1cm}
\begin{subfigure}{0.16\textwidth}
\begin{overpic}[width=\linewidth]{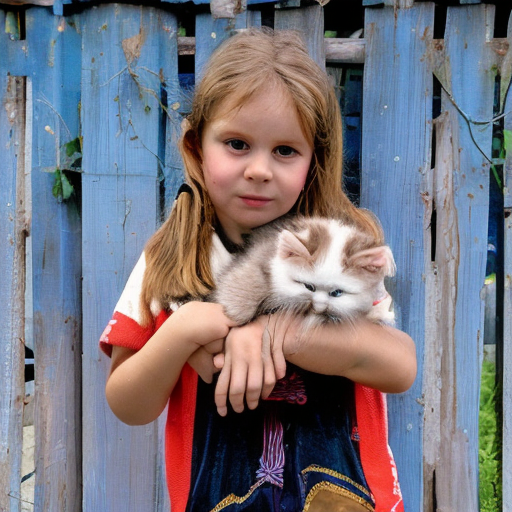}
\put(30,105){\small{Original}} 
\put(-18,40){\rotatebox{90}{\small{SG}}} 
\end{overpic}
\end{subfigure}
\hfill
\begin{subfigure}{0.16\textwidth}
\begin{overpic}[width=\linewidth]{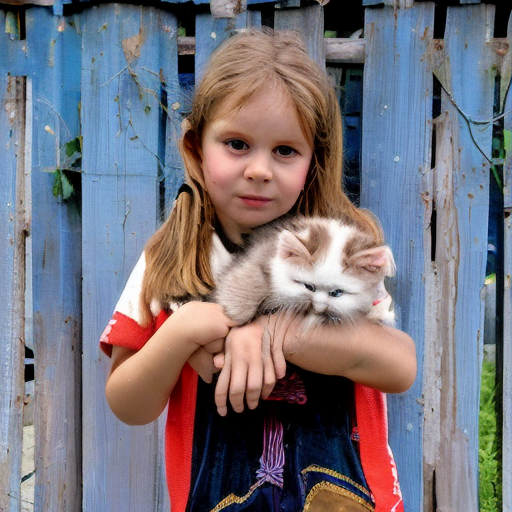}
\put(20,105){\small{Watermarked}}
\end{overpic}
\end{subfigure}
\hfill
\begin{subfigure}{0.16\textwidth}
\begin{overpic}[width=\linewidth]{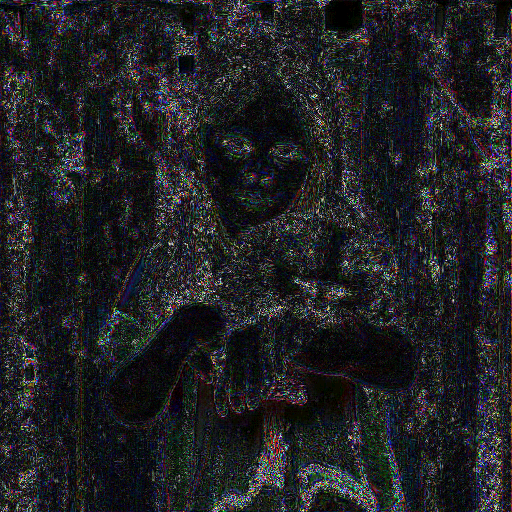}
\put(15, 105){\small{Pixel-wise ($\times$10)}}
\end{overpic}
\end{subfigure}
\hfill
\begin{subfigure}{0.16\textwidth}
\begin{overpic}
    [width=\linewidth]{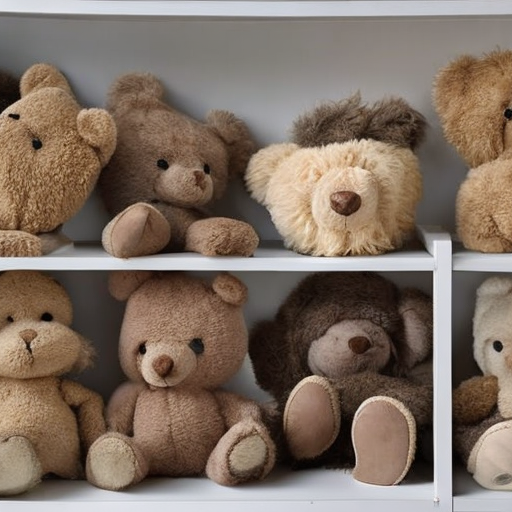}
\put(30,105){\small{Original}}
\end{overpic}
\end{subfigure}
\hfill
\begin{subfigure}{0.16\textwidth}
\begin{overpic}
[width=\linewidth]{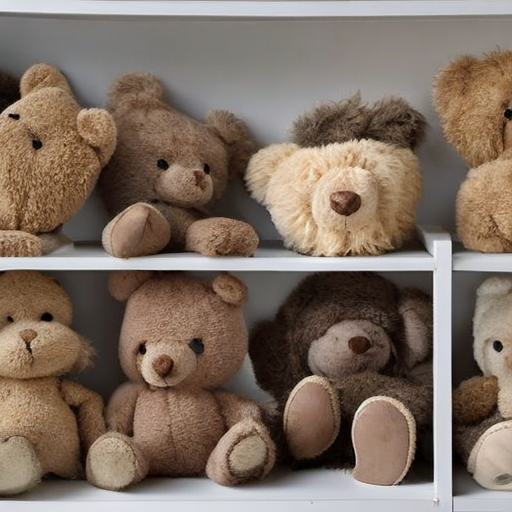}
\put(20,105){\small{Watermarked}}
\end{overpic}
\end{subfigure}
\hfill
\begin{subfigure}{0.16\textwidth}
\begin{overpic}
[width=\linewidth]{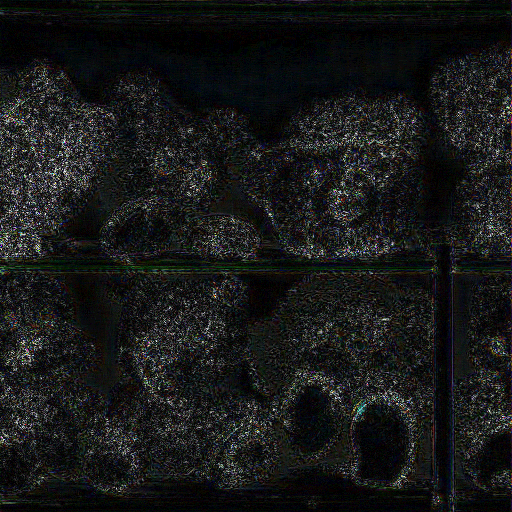}
\put(15, 105){\small{Pixel-wise ($\times$10)}}
\end{overpic}
\end{subfigure}

\begin{subfigure}{0.16\textwidth}
\begin{overpic}[width=\linewidth]{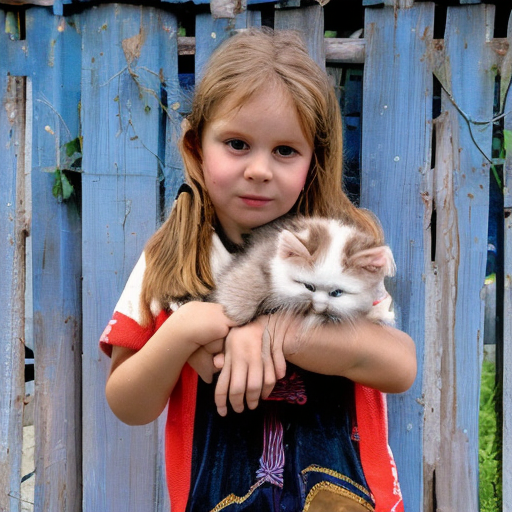}
\put(-18,25){\rotatebox{90}{\small{WOUAF}}}
\end{overpic}
\end{subfigure}
\hfill
\begin{subfigure}{0.16\textwidth}
\includegraphics[width=\linewidth]{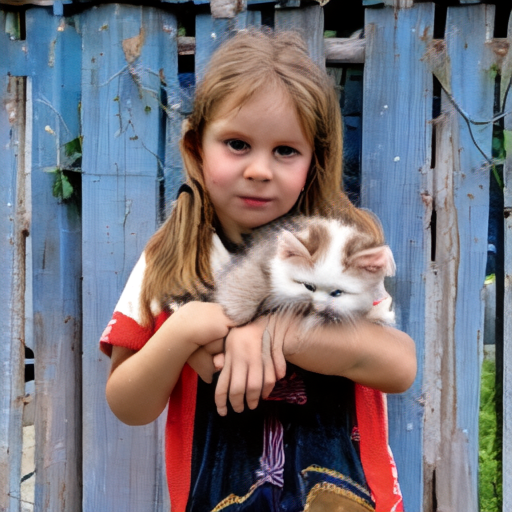}
\end{subfigure}
\hfill
\begin{subfigure}{0.16\textwidth}
\includegraphics[width=\linewidth]{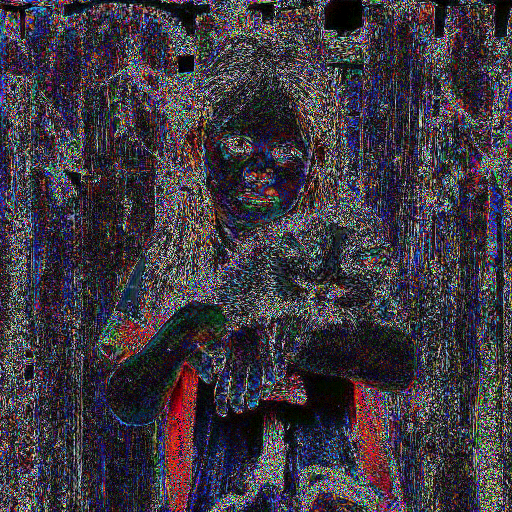}
\end{subfigure}
\hfill
\begin{subfigure}{0.16\textwidth}
\includegraphics[width=\linewidth]{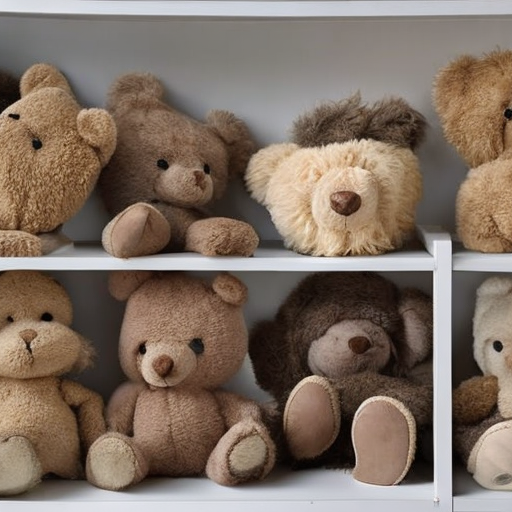}
\end{subfigure}
\hfill
\begin{subfigure}{0.16\textwidth}
\includegraphics[width=\linewidth]{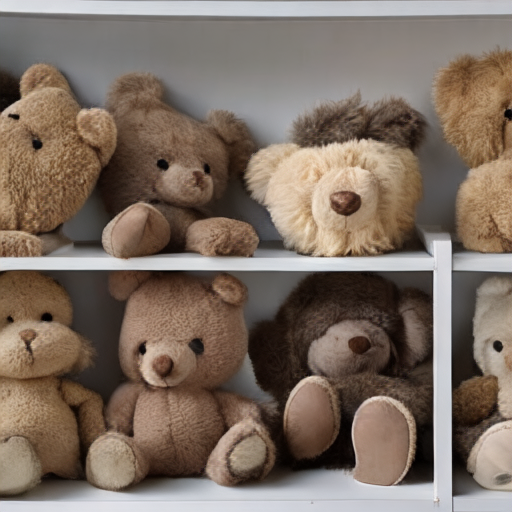}
\end{subfigure}
\hfill
\begin{subfigure}{0.16\textwidth}
\includegraphics[width=\linewidth]{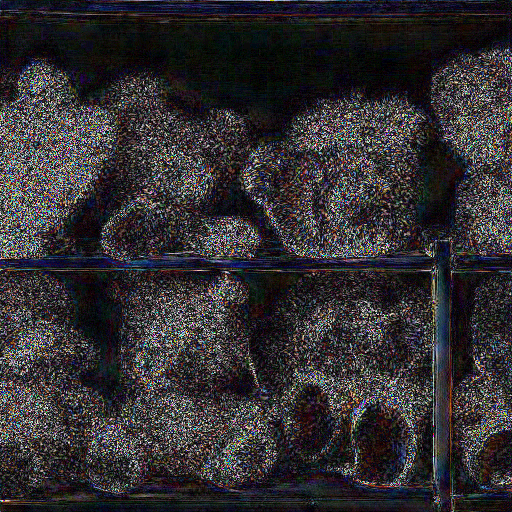}
\end{subfigure}

\begin{subfigure}{0.16\textwidth}
\begin{overpic}
[width=\linewidth]{sec/fig/00037_sd.png}
\put(-18,10){\rotatebox{90}{\small{TEAWIB (ours)}}}
\end{overpic}
\end{subfigure}
\hfill
\begin{subfigure}{0.16\textwidth}
\includegraphics[width=\linewidth]{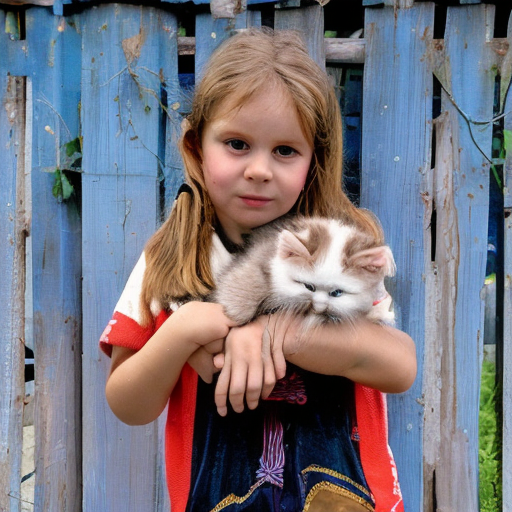}
\end{subfigure}
\hfill
\begin{subfigure}{0.16\textwidth}
\includegraphics[width=\linewidth]{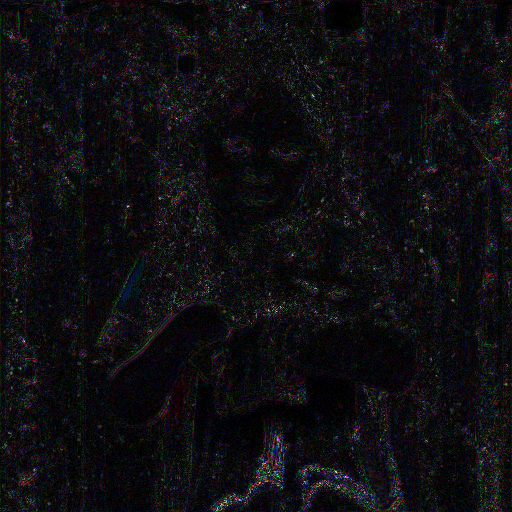}
\end{subfigure}
\hfill
\begin{subfigure}{0.16\textwidth}
\includegraphics[width=\linewidth]{sec/fig/sd_00941.png}
\end{subfigure}
\hfill
\begin{subfigure}{0.16\textwidth}
\includegraphics[width=\linewidth]{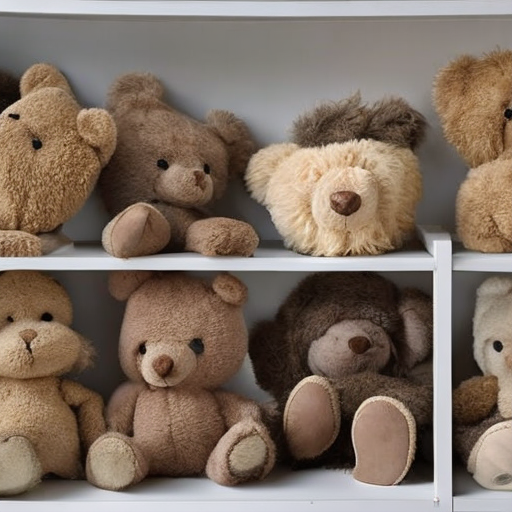}
\end{subfigure}
\hfill
\begin{subfigure}{0.16\textwidth}
\includegraphics[width=\linewidth]{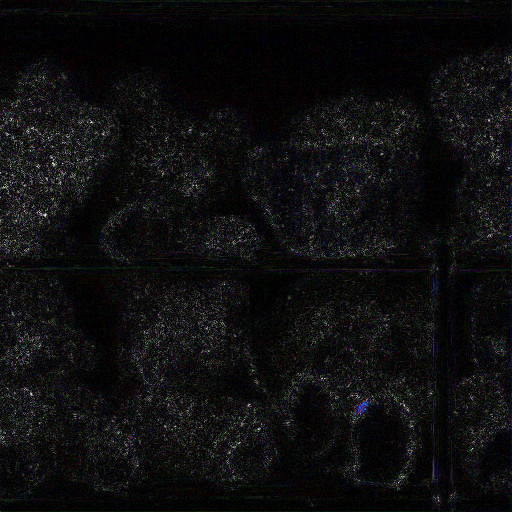}
\end{subfigure}
\caption{Qualitative comparison of TEAWIB with other ad-hoc watermark generation techniques on the MS-COCO validation set. Notably, our method preserves the high quality of the generated image and invisible watermark embedding.}
\label{fig:qualitative}
\end{figure*}

Tab.\ref{tab:quality-watermarking} shows the quantitative comparison between our TEAWIB and other merged-in generation methods. Remarkably, TEAWIB demonstrates superior performance across all metrics. The average PSNR has improved significantly from $30.0$~dB to approximately $39.2$~dB, the SSIM has increased from $0.89$ to $0.98$, and the lowest FID score is also achieved. Therefore, our TEAWIB can generate images that closely resemble those from the original SD models, in which the presence of watermarks is effectively imperceptible. We also provide the qualitative comparison in Fig.\ref{fig:qualitative}. It can be seen that the observed differences between original and watermarked images are negligible, and even when magnified tenfold, the pixel-wise discrepancies remain minimal.

\begin{table*}
[tbp]
    \centering
        \footnotesize
        \vspace{0.1cm} 
        \caption{Robustness analysis of TEAWIB and compared methods under various post-processing techniques.}\label{tab:robustness}
        \begin{adjustbox}{max width=\textwidth}
        \begin{tabular}{c *{8}{c}}
            \toprule
            \multirow{2}{*}{Model} & \multicolumn{8}{c}{Bit accuracy under various post-processing techniques} \\
            \cmidrule(lr){2-9}
            & Bright. $1.5$& Sharp $2.0$ & Sharp $1.5$ &  Text Overlay & Cont. $1.5$  & Crop $0.1$ & Sat. $2.0$ & Sat. $1.5$\\
            \midrule
            Stable Signature \cite{ref21} &0.9852 & 0.9829  & 0.9910& 0.9905 & 0.9808 & \textbf{0.9568}  & 0.9889&0.9910 \\
            WOUAF \cite{ref22} & 0.8403& 0.9758 & 0.9871& 0.9517 &  0.9588 & 0.5599 &0.8041&0.9607\\
            TEAWIB (Ours) & \textbf{0.9853}& \textbf{0.9906} & \textbf{0.9914}& \textbf{0.9914} & \textbf{0.9893} &  0.9295 &  \textbf{0.9898}&\textbf{0.9917} \\
            \bottomrule
        \end{tabular}
        \end{adjustbox}
    \hfill
    
\end{table*}

\subsubsection{Watermark Detection.}
\begin{figure*}[h]
    \begin{minipage}{0.24\linewidth}
        \centering
        \includegraphics[width=\linewidth, trim={0 0 0 0}, clip]{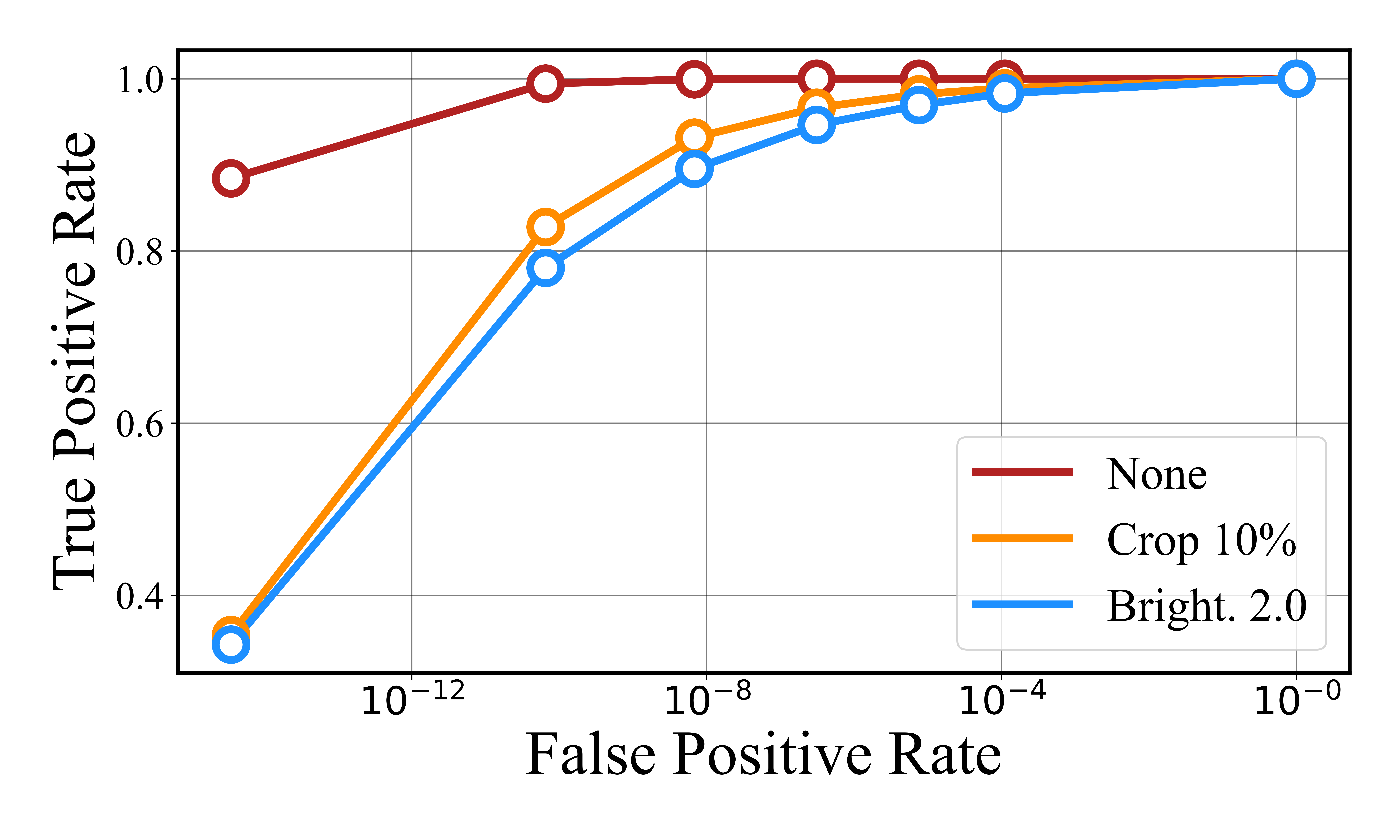}
        \caption{Model detection.}
        \label{fig:wtmk detection}
    \end{minipage}\hfill
    \begin{minipage}{0.24\linewidth}
        \centering
        \includegraphics[width=\linewidth, trim={0 0 0 0}, clip]{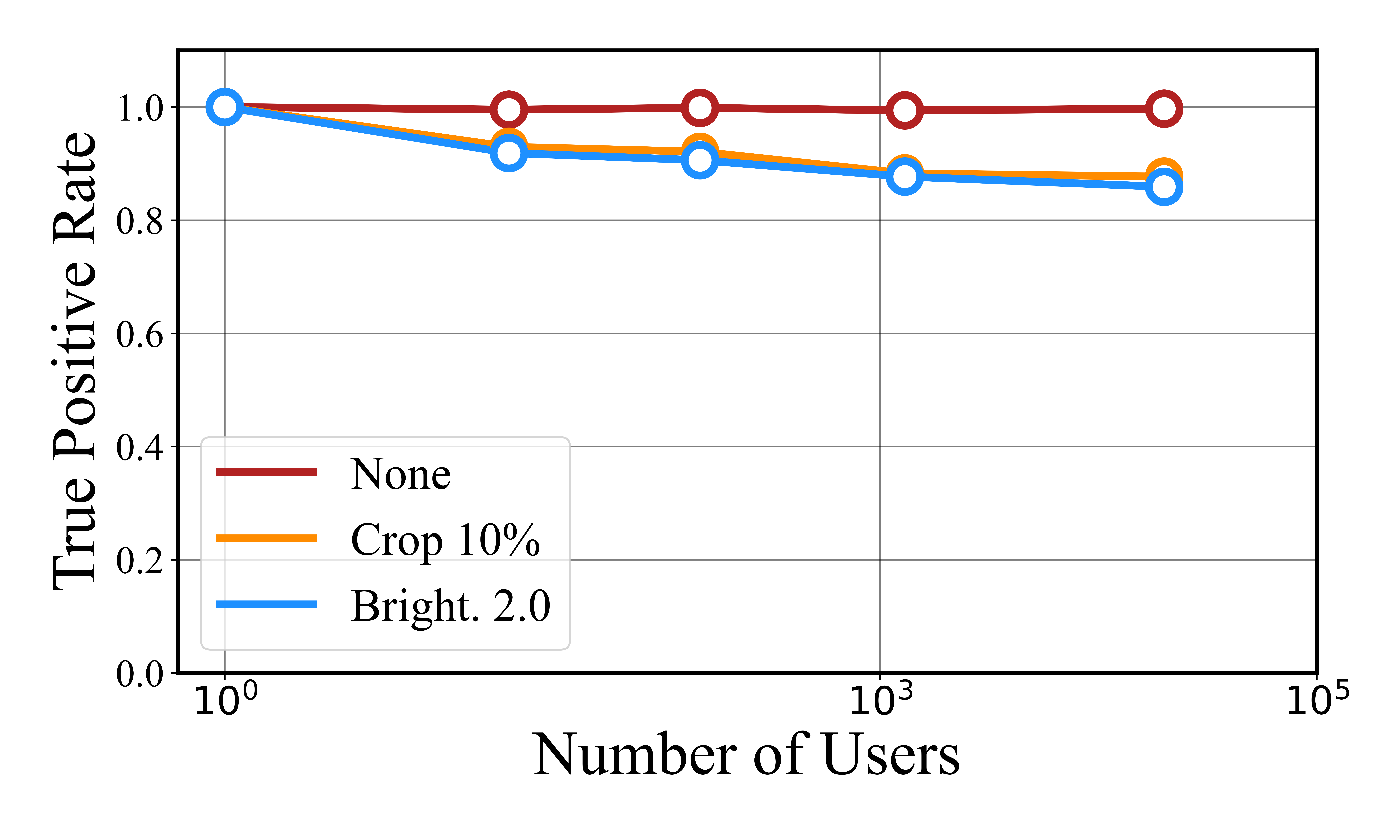}
        \caption{Model identification.}
        \label{fig:wtmk identification}
    \end{minipage}\hfill
    \begin{minipage}{0.24\linewidth}
        \centering
        \includegraphics[width=\linewidth, trim={0 0 0 0}, clip]{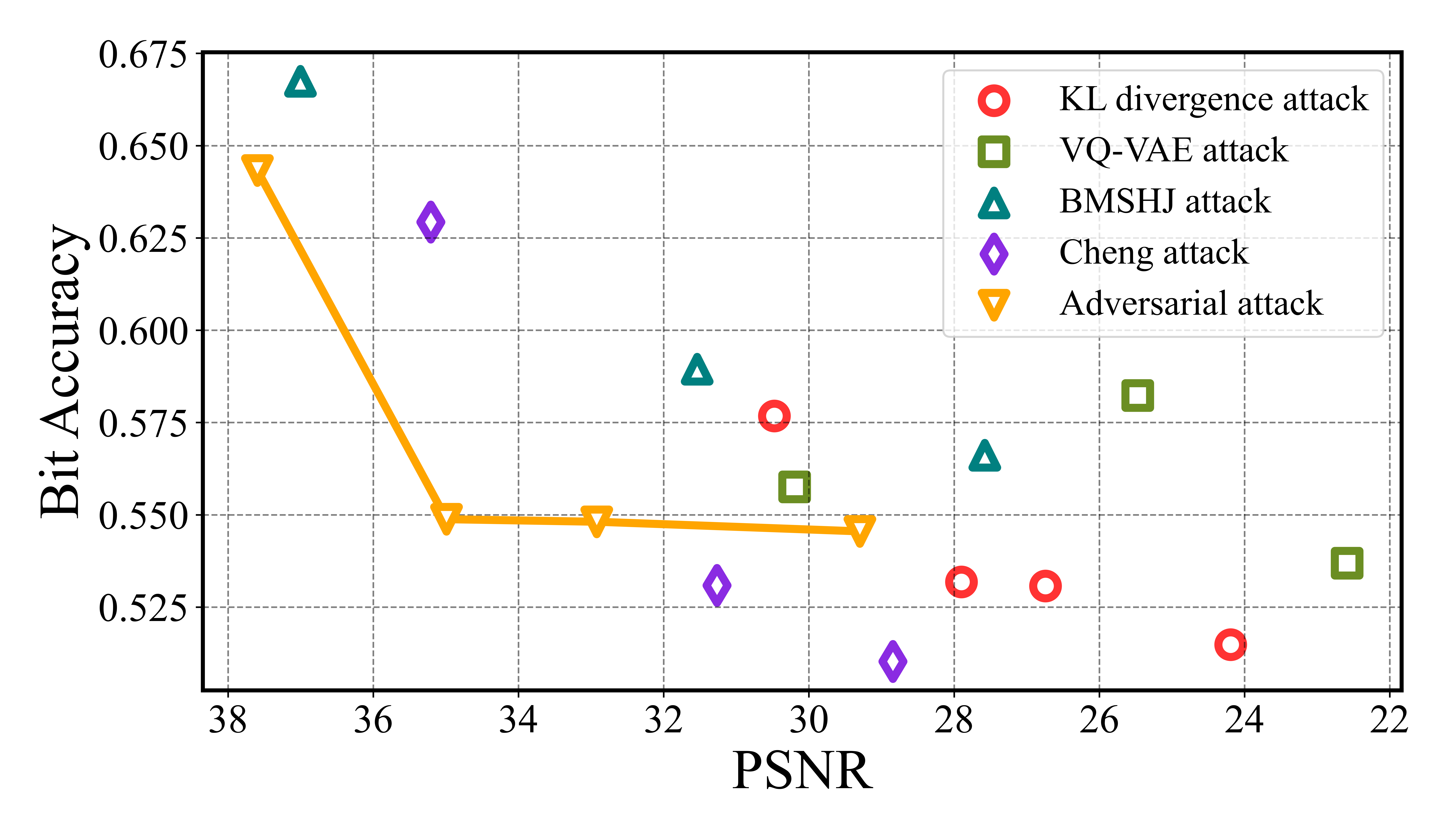}
        \caption{Image-level attacks.}
        \label{fig:autoencoder-wtmk}
    \end{minipage}\hfill
    \begin{minipage}{0.24\linewidth}
        \centering
        \includegraphics[width=\linewidth, trim={0 0 0 0}, clip]{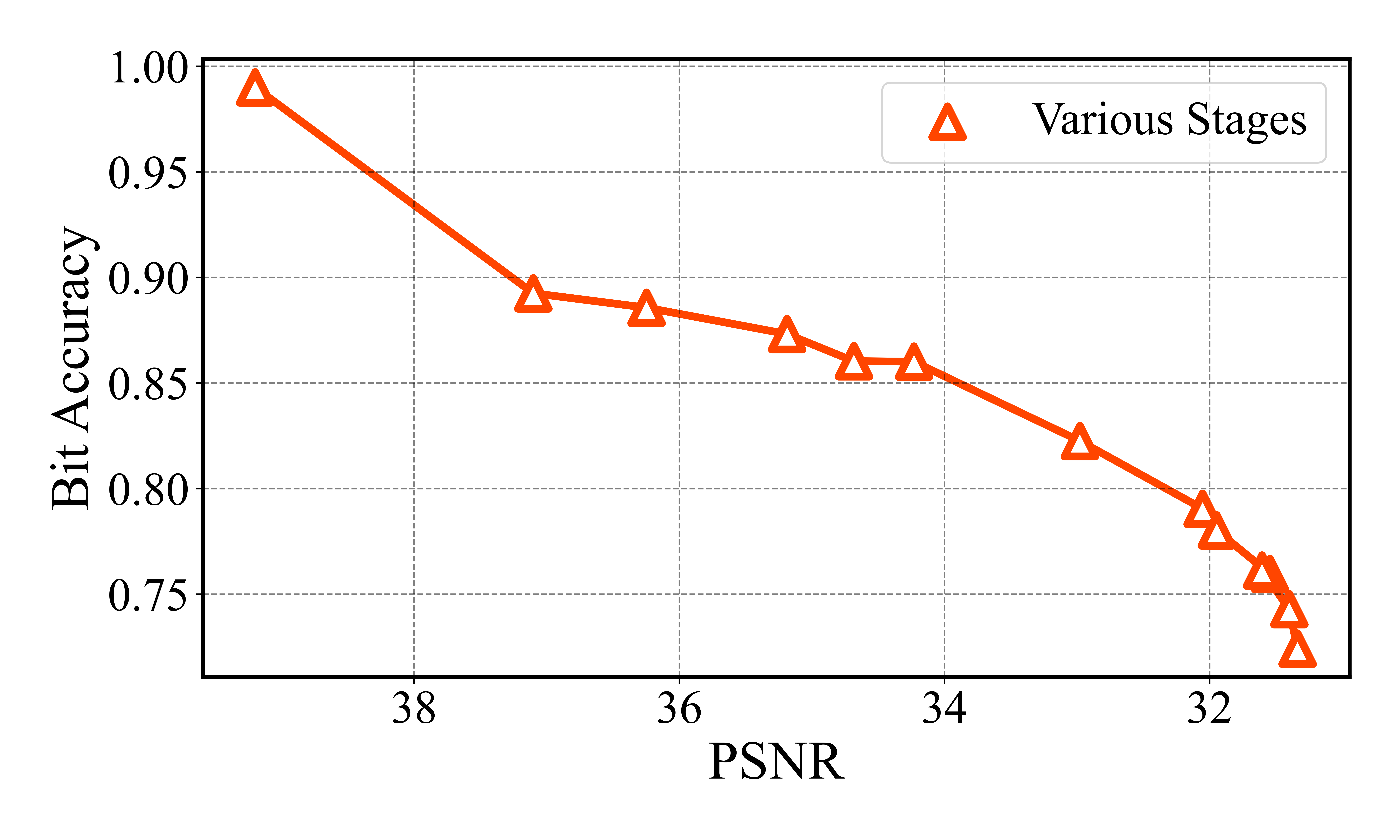}
        \caption{Model purification.}
        \label{fig:model purification}
    \end{minipage}\hfill
\end{figure*}

To evaluate the watermark detection performance, we use $10$ distinct watermarks and generate $1,000$ images for each watermark. Experimental results of watermark detection are provided in this subsection, where the detection criterion is defined by the matching bits $M(m,m')$:
\begin{equation}\label{eq: matching bits}
    M\left(m,m'\right) \geq \tau \,\,\textrm{ where }\,\, \tau\in 
\{0,\ldots,d_w\}.
\end{equation}
Here $\tau$ represents the manually selected threshold for detection. Formally, we test the statistical hypothesis $H_1$: ``$x$ was generated by Alice's model" against the null hypothesis $H_0$: ``$x$ was not generated by Alice's model". Under $H_0$ (\textit{i.e.} for vanilla images), we assume that bits $m'_1,\ldots, m'_k$ are independent and identically distributed (i.i.d.) Bernoulli random variables with parameter $0.5$.
Consequently, $M(m, m')$ follows a binomial distribution with parameters ($k$, $0.5$).
The False Positive Rate (FPR) is the probability that $M(m, m')$ exceeds the threshold $\tau$.
It is derived from the Cumulative Distribution Function (CDF) of the binomial distribution, and a closed-form expression can be written using the regularized incomplete beta function $I_x(a, b)$:
\begin{align}\label{eq:p-value}
    \text{FPR}(\tau) & = \mathbb{P}\left(M > \tau | H_0\right) = I_{0.5}(\tau+1, k - \tau),
\end{align}
where $x=0.5$. We select $10$ fixed random watermarks and generate $1,000$ images for each. We then report the averaged trade-off between the True Positive Rate (TPR) and the FPR, while varying $\tau\in \{0, .. ,48\}$. The TPR is measured directly, while the FPR is calculated by Eq. \ref{eq:p-value}. The results, displayed in Fig. \ref{fig:wtmk detection}, indicate that our method can achieve a detection accuracy of $99\%$ with a FPR of $10^{-11}$ if the image is free of any post-processing. However, we also observe that post-processing techniques can lead to a decrease in detection accuracy, which will be discussed later.

\subsubsection{Watermark Identification.}
Watermark identification refers to the scenario where the model owner extracts the watermark $m'$ from suspicious images and identifies the user who generated them. Suppose there are N candidates $m^{(1)}, m^{(2)}, \cdots, m^{(N)}$. In the identification task, the global FPR with respect to $\tau$ is defined as:
\begin{equation}\label{eq:globalFPR}
    \text{FPR}(\tau,N) = 1-(1-\text{FPR}(\tau))^N\approx N\cdot\text{FPR}(\tau).
\end{equation}
For evaluation, we randomly select $N'=1,000$ watermarks, each of which is used to generate $100$ images. For these generated images, we extract watermarks and compute the matching bits with all $N$ watermarks, selecting the user with the highest matching score. An image is predicted to be generated by that user if this score exceeds the threshold $\tau$. We adjust $\tau$ in Eq. \ref{eq:globalFPR} to fix the $\textrm{FPR}$ at $10^{-6}$. 

To evaluate the identification accuracy, we increase the total number of watermarks $N$ ($\gg N'$) by adding irrelevant watermarks, thereby testing the robustness of the identification process. The identification results are shown in Fig. \ref{fig:wtmk identification}. It can be seen that the identification accuracy of TEAWIB achieves 99\% when $N=20,000$, validating its capability of watermark identification.

\subsubsection{Watermark Robustness.}
In this paper, the robustness of the watermark is evaluated by adding various post-processing techniques before decoding the watermark. For each transformation, we generate $1,000$ images using the prompts from the MS-COCO validation set, each with a randomly selected watermark message. The average bit accuracy is reported in Tab. \ref{tab:quality-watermarking}. Compared to WOUAF, our TEAWIB demonstrates superior bit accuracy in the presence of post-processing. While Stable Signature supports only one designated watermark without retraining, making it easier to achieve high bit accuracy, TEAWIB still outperforms it under most post-processing techniques.


Further robustness analysis of TEAWIB and compared methods can be found in Tab. \ref{tab:robustness}, where ``Bright. $1.5$'' refers to increasing the image brightness by $1.5 \times$; ``Sharp $2.0$'' and ``Sharp $1.5$'' represent $2 \times$ and $1.5 \times$ the original image sharpness; ``Text Overlay'' involves adding texts at random positions in the image; ``Cont. $1.5$'' adjusts the image contrast to $1.5 \times$ the original value; ``Crop $0.1$'' reduces the image size to 10\% of the original; ``Sat. $2.0$'' and ``Sat. $1.5$'' refer to $2 \times$ and $1.5 \times$ the original saturation. Even though the watermark decoder is trained jointly during the training phase \textit{without any post-processing augmentation}, our method achieves an average accuracy over 98\%, demonstrating its robustness against common image post-processing techniques.

\begin{table*}[htbp]
\centering
    \renewcommand\arraystretch{1.10}
    \renewcommand\tabcolsep{9pt}
    \footnotesize
    \caption{Ablation study of TEAWIB. Each variant is evaluated based on four key metrics: PSNR, SSIM, $L_{\infty}$, and LPIPS. $\uparrow$ indicates the higher the better, $\downarrow$ indicates the lower the better. \textit{Abbreviation:} \underline{DWB}: \textbf{Dynamic Watermark Blending}; \underline{Noise}: \textbf{Noise Operation}; \underline{Aug}: \textbf{Augmentation Operation}; \underline{Joint}: \textbf{Jointly training watermark extractor}; \underline{LPIPS}: \textbf{Use LPIPS Loss}. \underline{ALL}: \textbf{Use Watermark-Informed Blending across all layers}.}

     \centering
    \vspace{-10pt}
    \tiny
    \resizebox{\linewidth}{!}{\begin{tabular}{l|cccccc|c|c|c|c}
    \hline
        \multicolumn{7}{c|}{\textbf{Training Components}} & \multicolumn{4}{c}{\textbf{Evaluation Metrics}} \\ \hline
        Variant & DWB &  Noise & Aug & Joint & LPIPS & ALL & PSNR$\uparrow$ & SSIM$\uparrow$ & $L_{\infty}$$\downarrow$ & LPIPS$\downarrow$ \\ \hline
        \multicolumn{7}{c|}{\textit{Reference Results of Baseline Model}}& \textcolor{gray}{28.4} & \textcolor{gray}{0.86} & \textcolor{gray}{123.5} & \textcolor{gray}{0.0614} \\ \hdashline
        {1} & \cmark &  & & & & \cmark & 34.8 & 0.97 & 64.8 &  0.0080 \\ \hdashline
        {2} & \cmark & \cmark & & \cmark & \cmark & \cmark & 34.8 & 0.97 &  65.5 & 0.0080 \\
        {3} & \cmark &  & \cmark & \cmark & \cmark & \cmark & 38.9 & 0.98 & 58.6 & 0.0051 \\ 
        {4} & \cmark & \cmark & \cmark & \cmark &  & \cmark & 38.9 & 0.98 & 59.6 & 0.0050 \\ 
        {5} & \cmark & \cmark & \cmark &  & \cmark & \cmark & 32.3 & 0.93 & 94.7 & 0.0357 \\
        {6} & \cmark & \cmark & \cmark & \cmark & \cmark &  & 31.0 & 0.91 & 100.963 & 0.0345 \\
        \hline
        {TEAWIB} & \cmark & \cmark & \cmark & \cmark & \cmark & \cmark & \textbf{39.2} & \textbf{0.98} & \textbf{57.8} & \textbf{0.0047} \\ \hline
    \end{tabular}}
    \label{tab:quality variants}
    \vspace{3pt}
\end{table*}



\subsection{Ablation Study}\label{model variants ablation}
\subsubsection{Watermark-Informed Blending.} 
The proposed WIB is a critical component of our framework, consisting of two main modules: Dynamic Watermark Blending (DWB) and Image Quality Preservation (IQP). DWB focuses on seamlessly integrating watermark-specific weights into the model, while IQP is designed to maintain high image quality through noise operation and augmentation operation.

Therefore, to thoroughly validate its effectiveness, we evaluated four variants to demonstrate the impact of these components on model performance, as detailed in Tab. \ref{tab:quality variants}: 1) the baseline model that uses only weight modulation for watermark insertion; 2) Variant 1 that uses only the DWB component, which significantly enhances the model's ability to produce higher-quality images; 3) Variant 2 excludes the augmentation operation in IQP, demonstrating a significant reduction in model efficacy and highlighting its crucial role in preserving robustness; 4) Variant 3 removes the noise operation in IQP, leading to a noticeable drop in performance and underscoring the necessity of this element for maintaining image quality.


\subsubsection{Training and Network Settings.} 
Effects of other components within the watermark-informed blending method are further investigated. Tab. \ref{tab:quality variants} also outlines various settings of TEAWIB, highlighting the importance of each component in our framework.

Key observations from results include: 1) Variant 4 excludes the LPIPS loss during training, resulting in a decline in performance. This reflects its importance in optimizing the perceptual quality of the output; 2) Variant 5 freezes the watermark decoder throughout the training process. This severely hinders the model's effectiveness, as evidenced by a marked decrease in performance; 3) Variant 6 omits the watermark-informed blending in the initial and final convolutional layers. This omission greatly compromises the image quality, further demonstrating the importance of these layers in the blending process. Our full model achieves the highest performance across all metrics, highlighting the effectiveness and synergy of the complete design.

\subsection{Deliberate Attack on Watermarks}
This subsection examines the robustness of our method against deliberate attempts to remove watermarks. We consider two types of attacks: image-level attacks, which operate directly on the image, and model-level attacks, which target the decoder $\mathcal{D}$. 
\subsubsection{Image Level Attacks.}
Auto-encoder attack and adversarial attack are common adopted methods to modify output images. The auto-encoder attack refers to passing the generated images through various auto-encoder models. In this paper, we use the VQ-VAE attack \cite{balle2018variational}, KL divergence attack \cite{ref3}, BMSHJ attack \cite{balle2018variational}, and Cheng attack \cite{cheng2020learned}. As shown in Fig. \ref{fig:autoencoder-wtmk}, the accuracy of watermark detection gradually decreases as the auto-encoder compression rate increases, eventually approaching a random guess level of around $50\%$. Although these attacks can reduce watermark detection accuracy, they also severely degrade image quality, making them impractical for the real-world usage.

However, adversarial attack becomes a concern when the watermark extractor $\mathcal{D}_w$ is leaked, attackers can employ adversarial techniques to replace the original watermark with a random one while preserving image quality. These attacks aim to minimize the $\ell_2$ distance between a pre-sampled random binary message and the decoded output, effectively substituting the original watermark. As shown in Fig. \ref{fig:autoencoder-wtmk}, such attacks can successfully remove watermarks. Therefore, it is necessary for the model owner to ensure the security of the watermark extractor.


\subsubsection{Model Level Attacks.}\label{sec: model_attack}
In the scenario where an attacker becomes aware of the presence of invisible watermarks, they may attempt to remove them by fine-tuning the diffusion decoder $\mathcal{D}$.
In this context, the attacker’s primary objective is to minimize the reconstruction error between $I_w$ and $I_o$. The PSNR values between the watermarked and purified images at various stages of fine-tuning are shown in Fig. \ref{fig:model purification}. Our observation indicates that attempts to remove watermarks through model purification result in a degradation of image quality. Therefore, significantly reducing bit accuracy without compromising image quality is challenging, as artifacts tend to emerge during the purification process.
\subsubsection{Model Collusion}
Model collusion attack is a scenario where users combine watermarked models to generate a colluded model, attempting to evade detection. We observed that the bit at position $i$ output by the watermark extractor remains identical if the corresponding bits of two users (e.g., John and Sam) are the same. However, if the bits differ at position $i$, the output becomes random.

\section{Conclusion}
In this paper, we propose a scalable watermarking method, TEAWIB, which enables effective user attribution for latent diffusion models through a novel watermark-informed blending strategy. TEAWIB excels at preserving the high quality of generated images while achieving near-perfect watermark extraction accuracy, making it robust for tracking content ownership. Additionally, we validate the robustness of TEAWIB against common post-processing techniques, ensuring resilience to typical image modifications.
\noindent\textbf{Limitation.} Currently, TEAWIB is limited to text-to-image generation. However, as TEAWIB requires only minimal adjustments to the Stable Diffusion decoder, we anticipate its adaptability to tasks such as image-to-image and video generation. Future work will focus on broadening TEAWIB’s application scope.



{
    \small
    \bibliographystyle{ieeenat_fullname}
    \bibliography{main}

}


\end{document}